\documentclass[aps,prb,twocolumn,superscriptaddress,showpacs]{revtex4-1}

\usepackage{graphicx}
\usepackage{dcolumn}
\usepackage{amsthm}
\usepackage{amsmath}
\usepackage{bm}
\usepackage{longtable}
\usepackage{color}
 
\begin{document}

\title{Disentangling surface and bulk transport in topological-insulator $p$-$n$ junctions}

\author{Dirk Backes}
\email{db639@cam.ac.uk}
\affiliation{Cavendish Laboratory, University of Cambridge, J. J. Thomson Avenue, Cambridge CB3 0HE, United Kingdom}
\author{Danhong Huang}
\affiliation{Air Force Research Laboratory, Space Vehicles Directorate, Kirtland Air Force Base, New Mexico 87117, USA}
\author{Rhodri Mansell}
\affiliation{Cavendish Laboratory, University of Cambridge, J. J. Thomson Avenue, Cambridge CB3 0HE, United Kingdom}
\author{Martin Lanius}
\affiliation{Peter Gr\"unberg Institute (PGI-9), Forschungszentrum J\"ulich, 52425 J\"ulich, Germany}
\author{J\"orn Kampmeier}
\affiliation{Peter Gr\"unberg Institute (PGI-9), Forschungszentrum J\"ulich, 52425 J\"ulich, Germany}
\author{David Ritchie}
\affiliation{Cavendish Laboratory, University of Cambridge, J. J. Thomson Avenue, Cambridge CB3 0HE, United Kingdom}
\author{Gregor Mussler}
\affiliation{Peter Gr\"unberg Institute (PGI-9), Forschungszentrum J\"ulich, 52425 J\"ulich, Germany}
\author{Godfrey Gumbs}
\affiliation{Department of Physics and Astronomy, Hunter College of the City University of New York, 695 Park Avenue, New York, New York 10065, USA}
\author{Detlev Gr\"utzmacher}
\affiliation{Peter Gr\"unberg Institute (PGI-9), Forschungszentrum J\"ulich, 52425 J\"ulich, Germany}
\author{Vijay Narayan}
\email{vn237@cam.ac.uk}
\affiliation{Cavendish Laboratory, University of Cambridge, J. J. Thomson Avenue, Cambridge CB3 0HE, United Kingdom}

\date{\today}

\begin{abstract}
By combining $n$-type $\mathrm{Bi_2Te_3}$ and $p$-type $\mathrm{Sb_2Te_3}$ topological insulators, vertically stacked  $p$-$n$ junctions can be formed, allowing to position the Fermi level into the bulk band gap and also tune between $n$- and $p$-type surface carriers. Here we use low-temperature magnetotransport measurements to probe the surface and bulk transport modes in a range of vertical $\mathrm{Bi_2Te_3/Sb_2Te_3}$ heterostructures with varying relative thicknesses of the top and bottom layers. With increasing thickness of the $\mathrm{Sb_2Te_3}$ layer we observe a change from $n$- to $p$-type behavior via a specific thickness where the Hall signal is immeasurable. Assuming that the the bulk and surface states contribute in parallel, we can calculate and reproduce the dependence of the Hall and longitudinal components of resistivity on the film thickness.  This highlights the role played by the bulk conduction channels which, importantly, cannot be probed using surface sensitive spectroscopic techniques. Our calculations are then buttressed by a semi-classical Boltzmann transport theory which rigorously shows the vanishing of the Hall signal. Our results provide crucial experimental and theoretical insights into the relative roles of the surface and bulk in the vertical topological $p$-$n$ junctions.
\end{abstract}


\maketitle

\section{Introduction}

Topological insulators (TIs) are bulk insulators with exotic ``topological surface states''~\cite{Hasan:2010} (TSS) which are robust to backscattering from non-magnetic impurities, exhibit spin-momentum locking\,\cite{Hsieh:2009b}, and have a Dirac-type dispersion\,\cite{Xia:2009,Chen:2009,Hsieh:2008}. These unique characteristics present several opportunities for applications in spintronics, thermoelectricity, and quantum computation. However, a major drawback of ``early generation'' TIs such as $\mathrm{Bi_{1-x}Sb_x}$~\cite{Hsieh:2008} and $\mathrm{Bi_2Se_3}$~\cite{Xia:2009, Hsieh:2009b} is that the Fermi level $E_{\mathrm{F}}$ intersects the conduction/valence bands, thus giving rise to finite conductivity in the bulk. This non-topological conduction channel conducts in parallel to the TSS and in turn subverts the overall topological nature. Thus, in order to create {\it bona fide} TIs, the Fermi level $E_{\mathrm{F}}$ needs to be tuned within the bulk band gap, and this has previously been achieved by means of electrical gating~\cite{Chen:2010, Chen2011, Checkelsky:2011, Steinberg:2011}, doping~\cite{Chen:2009, Kong:2011, Zhang:2011, Weyrich:2016}, or, as recently reported, by creating $p$-$n$ junctions from two different TI films~\cite{Zhang:2013, Eschbach:2015}. 

In Ref.~\onlinecite{Eschbach:2015} a ``vertical topological $p$-$n$ junction'' was realized by growing an $n$-type $\mathrm{Bi_2Te_3}$ layer capped by a layer of $p$-type $\mathrm{Sb_2Te_3}$, and it was shown that varying the relative layer thicknesses serves to tune $E_{\mathrm{F}}$ without the use of an external field. Importantly, such bilayer systems are expected to be significantly less disordered than doped materials such as $\mathrm{(Bi_{1-x}Sb_x)_2Te_3}$ in which inhomogeneity of the dopants is a constant problem~\cite{Weyrich:2016, Lanius:2016}. Furthermore, and in sharp contrast to doped TIs, the intrinsic $p$ and $n$ character of the individual layers presents remarkable opportunities towards the observation of novel physics including Klein tunneling~\cite{Klein:1929, Kastnelson:2006}, spin interference effects at the $p$-$n$ interface~\cite{Ilan:2015}, and topological exciton condensates~\cite{Seradjeh:2009}. However, currently there exists little understanding of the bulk conduction in such topological $p$-$n$ junctions, primarily because ARPES used in Ref.~\onlinecite{Eschbach:2015} is a surface-sensitive method. This is especially noteworthy in light of the fact that the band structure varies along the depth of the TI $p$-$n$ junction slab, in sharp contrast to the essentially constant band gap within the bulk of $\mathrm{(Bi_{1-x}Sb_x)_2Te_3}$-type compounds. Understanding and minimizing the bulk conduction channels in TI $p$-$n$ junctions is crucial in order to realize their technological potential as well as to gain access to the exotic physics they can host.

\section{Experiment}

$\mathrm{Bi_2Te_3/Sb_2Te_3}$-bilayers (BST) were grown on phosphorous doped Si substrates using molecular beam epitaxy (MBE). Details of the MBE sample preparation can be found in Ref.~\onlinecite{Eschbach:2015}. In all the samples, the bottom $\mathrm{Bi_2Te_3}$-layer had thickness $t_\mathrm{BiTe}$~=~6\,nm while the top $\mathrm{Sb_2Te_3}$-layers had thicknesses $t_\mathrm{SbTe} = $ 6.6\,nm (BST6), 7.5\,nm (BST7), 15\,nm (BST15), and 25\,nm (BST25), respectively. The layers were patterned into Hall bars of width $W = \mathrm{200\,\mu m}$ and length $L = \mathrm{1000\,\mu m}$ using photoresist as a mask for ion milling, and Ti/Au contact pads were deposited for electrical contact. Low-$T$ electrical measurements were carried out using lock-in techniques in a He-3 cryostat with a base temperature of 280\,mK and a 10\,T superconducting magnet. Both longitudinal ($R_\mathrm{xx}$) and transverse ($R_\mathrm{xy}$) components of resistance were measured.

\section{Results}

\begin{figure}
	\centering
	\includegraphics[width= \columnwidth]{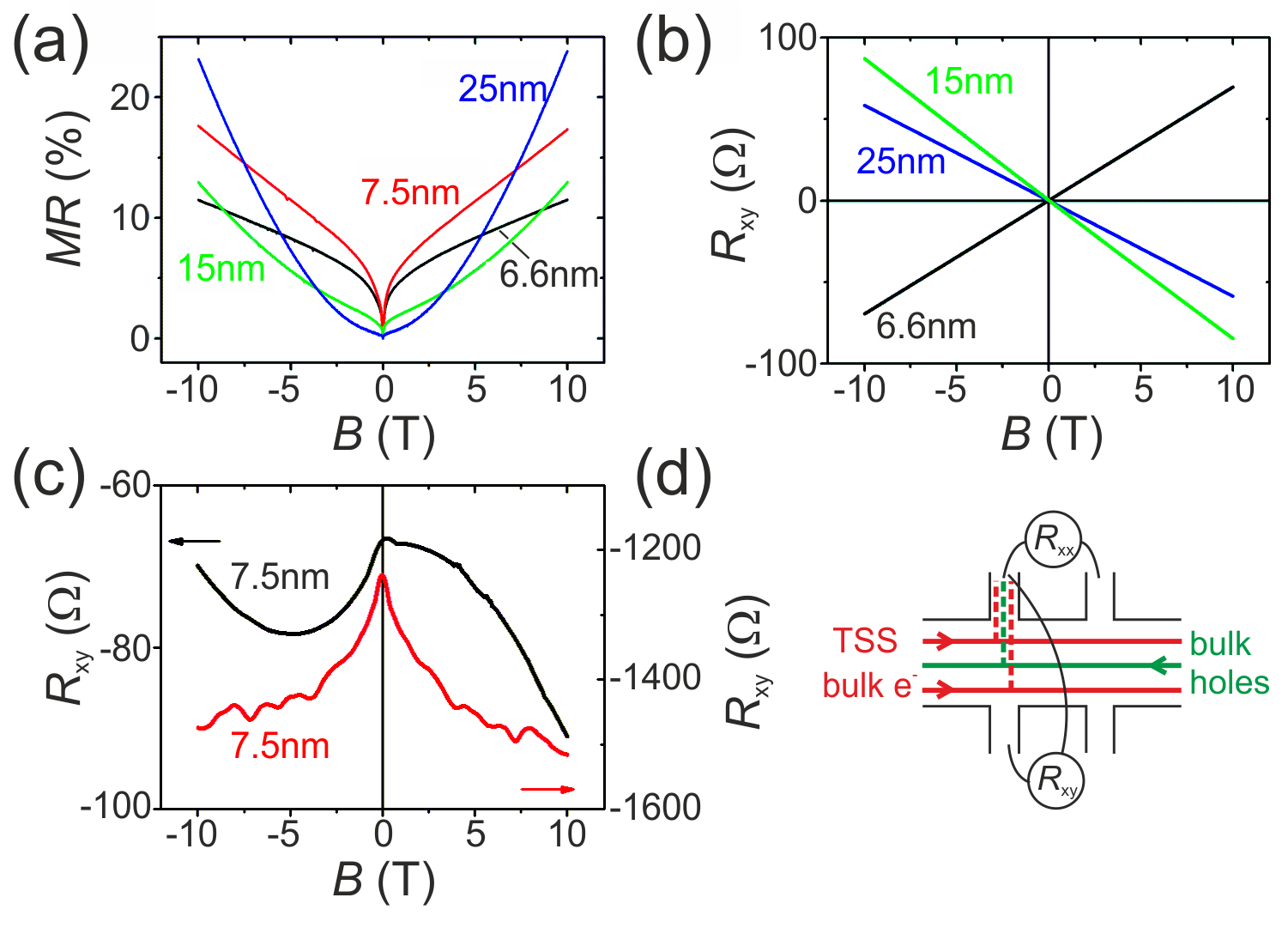} 
	\caption{(a) MR and (b), (c) $R_\mathrm{xy}$ as a function of $B$ for different $t_\mathrm{SbTe}$. All curves are measured at 280\,mK. The high field MR is linear for thin samples and changes to parabolic for thicker samples. Cusp-like deviations at low fields are due to WAL corrections. The sign change of the slope in (b) indicates transport by electrons for BST6 and by holes for BST15 and BST25. No Hall slope is visible in (c) for two different pairs of contacts of BST7. (d) The schematic shows the charge transport channels in a longitudinal and transverse measurement setup. Trajectories of TSS and bulk electrons are shown in red and of bulk holes in green.}
	\label{fig1}
\end{figure}

Figure~\ref{fig1}(a) shows the longitudinal magnetoresistance (MR) $\equiv [R_\mathrm{xx}(B)-R_\mathrm{xx}(0)]/R_\mathrm{xx}(0)$ of the various samples considered. We find that  above $\sim 2$\,T the MR in BST6 and BST7 is manifestly linear whereas the MR in BST15 and BST25 appears to be neither purely linear nor quadratic. While there is experimental evidence suggesting an association between linear MR and linearly dispersive media~\cite{Qu:2010,Wang:2012b,Liang:2015}, as well as a theoretical basis for this association~\cite{Abrikosov:1998}, we note that disorder can also render giant linear MR~\cite{Parish:2003,Narayanan:2015} by admixing longitudinal and Hall voltages. In Fig.~\ref{fig1}(b) we see that $R_\mathrm{xy}$ is linear in $B$ and its slope changes sign from positive (BST6) to negative (BST15 and BST25). This is simply a reflection of different charge carrier types of $\mathrm{Bi_2Te_3}$ ($n$-type) and $\mathrm{Sb_2Te_3}$ ($p$-type), where electrons (holes) dominate transport when $\mathrm{Sb_2Te_3}$ is thin (thick). Intriguingly, Fig.~\ref{fig1}(c) shows $R_\mathrm{xy}$ vs $B$ measured in two different Hall bar devices of BST7 to be strongly non-linear and non-monotonic. Qualitatively, it appears as if $R_\mathrm{xy}$ is picking up a large component of $R_\mathrm{xx}$ despite the Hall probes being aligned to each other with lithographic ($\mu$m-scale) precision. We conjecture, therefore, that BST7 is very close to where the Hall coefficient $R_\mathrm{H}$ precisely changes from positive to negative. Seemingly to the contrary, ARPES measurements in Ref.~\onlinecite{Eschbach:2015} reveal that $E_{\mathrm{F}}$ intersects the Dirac point in samples with 15\,nm~$< t_\mathrm{SbTe} <$~25\,nm, in which parameter regime Fig.~\ref{fig1}(b) indicates a net excess of $p$-type carriers. The investigation of this discrepancy is the major focus of this paper.

\begin{figure}
	\centering
	\includegraphics[width=\columnwidth]{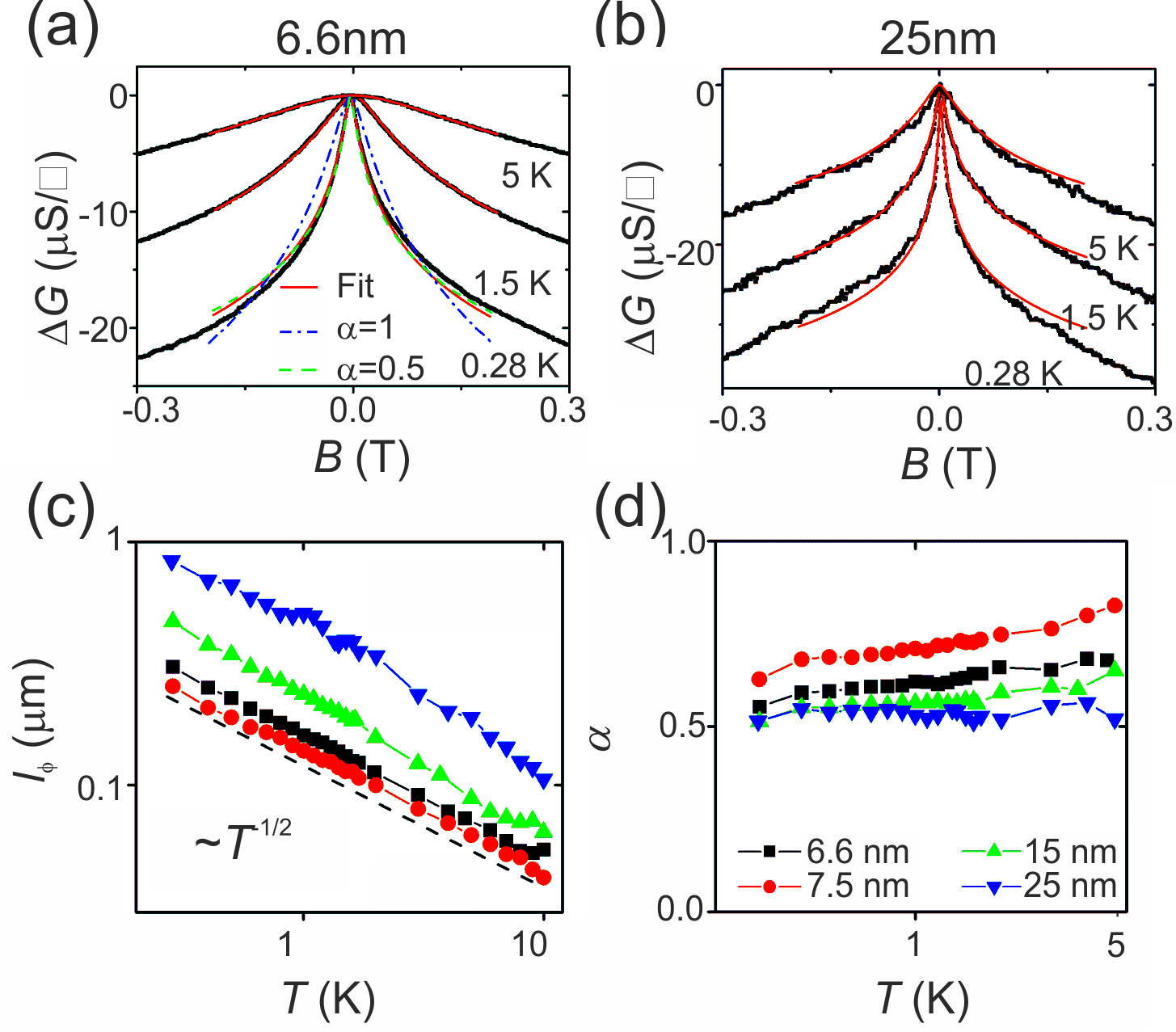}
	\caption{(a), (b) Weak antilocalization peaks for two different $\mathrm{Sb_2Te_3}$ thicknesses and at three different temperatures. Fits to the measurements, based on the HLN model, are shown in straight red lines, while curves with $\mathrm{\alpha}$ at 0.5 (green dashed line) and 1 (blue dashed-dotted line) allow to estimate the error. (c) $l_\mathrm{\phi}$ as a function of $T$ for various $t_\mathrm{SbTe}$ in a log-log plot. All curves are proportional to $\propto T^{-0.5}$ (dashed line) but shifted with respect to each other. (d) $\alpha$ as a function of $T$ for various $t_\mathrm{SbTe}$. }
	\label{fig2}
\end{figure}

Figures~\ref{fig2}(a) and (b) show the low-field MR where a pronounced ``weak anti-localisation'' (WAL) cusp is visible at zero magnetic field ($B$).  The WAL corrections are well-described by the model of Hikami, Larkin and Nagaoka (HLN)~\cite{Hikami:1980}

\begin{equation}
\begin{aligned}
\label{HLN}
\Delta \sigma_\mathrm{{xx}}^\mathrm{{2D}} &\equiv \sigma_\mathrm{{xx}}^\mathrm{{2D}}(B) - \sigma_\mathrm{{xx}}^\mathrm{{2D}}(0)\\ &= \alpha \frac{e^2}{2 \pi^2 \hbar}\left[ \ln \left( \frac{\hbar}{4eB l_\mathrm{\phi}^2}\right) - \psi \left( \frac{1}{2} + \frac{\hbar}{4eB l_\mathrm{\phi}^2}\right) \right].\\
\end{aligned}
\end{equation}

\noindent Here, $\sigma_\mathrm{{xx}} \equiv (L/W) R_\mathrm{xx}/(R_\mathrm{xx}^2 + R_\mathrm{xy}^2)$ and the superscript 2D indicates that the equation is valid for a two-dimensional conducting sheet, $\alpha$ is a parameter = 0.5 for each 2D WAL channel, $e$ is the electronic charge, $\hbar$ is Planck's constant divided by 2$\pi$, $l_\phi$ is the phase coherence length, and $\psi$ is the digamma function.

Figure~\ref{fig2}(c) shows the $T$-dependence of $l_\mathrm{\phi}$ for all samples. We find that $l_\mathrm{\phi} \propto T^{-\mathrm{p}/2}$, where the exponent $\mathrm{p} = 1$ is in line with 2D Nyquist scattering~\cite{Altshuler:1998, Takagaki:2012} due to electron-electron scattering processes. The second fitting parameter $\alpha$ is depicted in Fig.~\ref{fig2}(d) and we find values consistent with $\alpha = 0.5$ [error estimates on $\alpha$ can be found in Fig.\,\ref{fig2}(a) and a discussion in Appendix A]. This is consistent with several previous reports on TI thin films~\cite{Steinberg:2011, Garate:2012, Veldhorst:2013, Nguyen:2016}.

\section{Discussion}

\subsection{Three-channel model}

Having ascertained that the transport characteristics of the $\mathrm{Bi_2Te_3/Sb_2Te_3}$ heterostructures are consistent with conventional TI behavior, we now proceed to understand the Hall characteristics. It is well known that the TIs $\mathrm{Bi_2Te_3}$ and $\mathrm{Sb_2Te_3}$ show bulk conduction in addition to the TSS. Thus, we start with a simple picture of three independent conduction channels: bulk $n$- and $p$-type layers corresponding to the $\mathrm{Bi_2Te_3}$ and $\mathrm{Sb_2Te_3}$ layers, respectively, and a TSS on the top surface. While in principle a TSS exists also at the interface with the substrate, it is expected that its contribution to the conductivity is largely diminished due to the strongly disordered TI-substrate interface~\cite{Schubert:2012, Veldhorst:2013}. Thus as a first approximation, we do not consider the bottom TSS.

Our starting point is the expressions for $\sigma_\mathrm{xx}$ and $R_{\mathrm{H}}$ in a multi-channel system~\cite{Kittel:1986, Ren:2010, Eguchi:2016} 

\begin{equation}
\label{Long}
\sigma_\mathrm{xx}= e\,n_\mathrm{p}\mu_\mathrm{p}-e\,n_\mathrm{n}\mu_\mathrm{n}\pm e\,n_\mathrm{t}\mu_\mathrm{t}\\
\end{equation}

\begin{equation}
\label{Hall}
R_\mathrm{H}(t_\mathrm{SbTe})  \equiv \frac{1}{e \cdot n_\mathrm{eff}}=\frac{n_\mathrm{p} \mu_\mathrm{p}^2 -n_\mathrm{n} \mu_\mathrm{n}^2\pm n_\mathrm{t}(t_\mathrm{SbTe}) \mu_\mathrm{t}^2}{e(n_\mathrm{p} \mu_\mathrm{p}+n_\mathrm{n}\mu_\mathrm{n}+n_\mathrm{t}(t_\mathrm{SbTe}) \mu_\mathrm{t})^2}.
\end{equation}

\noindent Here $n_\mathrm{eff}$ is the effective carrier concentration, $e$ is the charge of an electron and $-e$ is the charge of a hole, the subscripts $n,p$ and $t$ signify bulk electrons, bulk holes, and surface carriers, respectively, $n_\mathrm{i}$ are carrier concentrations, and $\mu_\mathrm{i}$ represent the mobility of the charge carriers. The $\pm$ indicates, respectively, negative ($t_\mathrm{SbTe}<20$\,nm) and positive charge carriers ($t_\mathrm{SbTe}>20$\,nm) in the TSS. The following literature values for the bulk layers are assumed: $n_\mathrm{BiTe}=8\times10^{19}~\mathrm{cm}^{-3}$ and $\mu_\mathrm{n}=50\,\mathrm{cm^2V^{-1}s^{-1}}$ for $\mathrm{Bi_2Te_3}$~\cite{Weyrich:2016} and $n_\mathrm{SbTe}=4.5\times10^{19}~\mathrm{cm}^{-3}$ and $\mu_\mathrm{p}=300\,\mathrm{cm^2V^{-1}s^{-1}}$ for $\mathrm{Sb_2Te_3}$ ~\cite{Weyrich:2016, Takagaki:2012, Horak:1995}. In order to compare $n_\mathrm{BiTe}$ and $n_\mathrm{SbTe}$ to the TSS carrier concentration, we convert them to effective areal densities as $n_\mathrm{n} \equiv n_\mathrm{BiTe}\cdot t_\mathrm{BiTe}$ and $n_\mathrm{p} \equiv n_\mathrm{SbTe}\cdot t_\mathrm{SbTe}$. It can be shown that $n_t \propto E^2_\mathrm{B}$ where $E_\mathrm{B}$ is the difference between $E_\mathrm{F}$ and Dirac point [see Eq.\,(\ref{eqnb3}), Appendix B] and $E_\mathrm{B}$, in turn, can be retrieved from ARPES measurements in Ref.~\cite{Eschbach:2015}. $\mu_\mathrm{t}$ is used as a fitting parameter.

\begin{figure}
	\centering
	\includegraphics[width= \columnwidth]{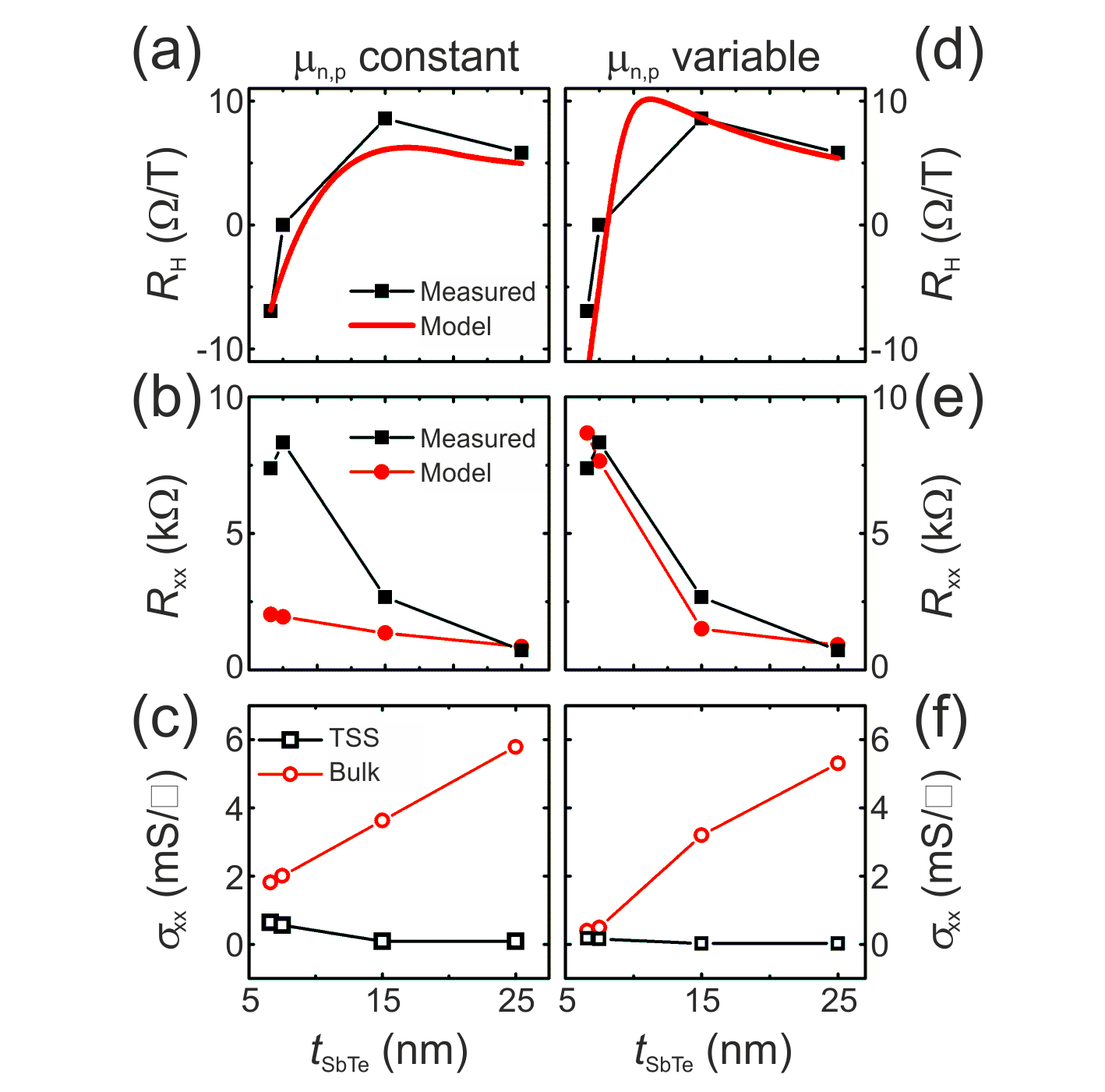}
	\caption{(a), (d) Hall slopes $R_\mathrm{H}$ determined from the Hall measurements in Fig.~\ref{fig1}(b) (black square), and fitted using Eq.~(\ref{Hall}) (red lines). The bulk mobilities $\mu_\mathrm{n,p}$ were kept constant in (a) and reduced for low thicknesses in (d).  (b), (c) Comparison of measured (black squares) and calculated total resistance (red disks), and conductivity of the TSS (black open squares) and of the bulk (red open disks), using fitting parameters from (a). (e), (f) Same as (b) and (c) but using fitting parameter from (d). All variables are a function of $t_\mathrm{SbTe}$.}
	\label{fig3}
\end{figure}

Figure~\ref{fig3}(a) shows $R_\mathrm{H}$ as predicted by the model using the above parameters to be in good agreement with the measured values. However, for the same parameters we find that $R_\mathrm{xx} \equiv (L/W)\sigma_\mathrm{xx}$ is significantly underestimated especially for low $t_\mathrm{SbTe}$ [see Fig.~\ref{fig3}(b)]. A likely source of this discrepancy is that the bulk $\mu_\mathrm{i}$ values are not applicable for the ultra-thin films. This is especially so considering the fact that a depletion zone will form at the $p$-$n$ interface. Determining the exact profile of the charge carrier density at the interface is beyond the scope of this paper and instead, we demonstrate that an \textit{ad hoc} thickness-dependent reduction of $\mu_\mathrm{i}$ of the \textit{bulk} layers with all other parameters unchanged, can significantly improve the quality of the predictions. Figure~\ref{fig3}(d) shows the result of a fit in which $\mu_\mathrm{p}$ and $\mu_\mathrm{n}$ are reduced to 20\% of their bulk value in BST6 and BST7, and to 95\% of their bulk value in BST15 and BST25. Not only do we obtain excellent agreement with the $R_\mathrm{H}$ data, the model is also able to accurately predict $R_\mathrm{xx}$ [see Fig.~\ref{fig3}(e)]. The obtained value of $\mu_\mathrm{t}=281\pm 17$\,$\mathrm{cm^2V^{-1}s^{-1}}$ is well within the range of previous studies in ultra-thin TIs where the TSS dominate transport~\cite{Zhang:2011}.

Figure~\ref{fig3}(f) shows the important physical insight we arrive at on the basis of this simple model: the bulk contribution is drastically reduced in thin films [see Fig.~\ref{fig3}(c)], with the TSS eventually dominating the overall conductivity $\sigma_\mathrm{tot}$ [see Fig.~\ref{fig3}(f)]. 

\begin{figure}
	\centering
	\includegraphics[width= \columnwidth]{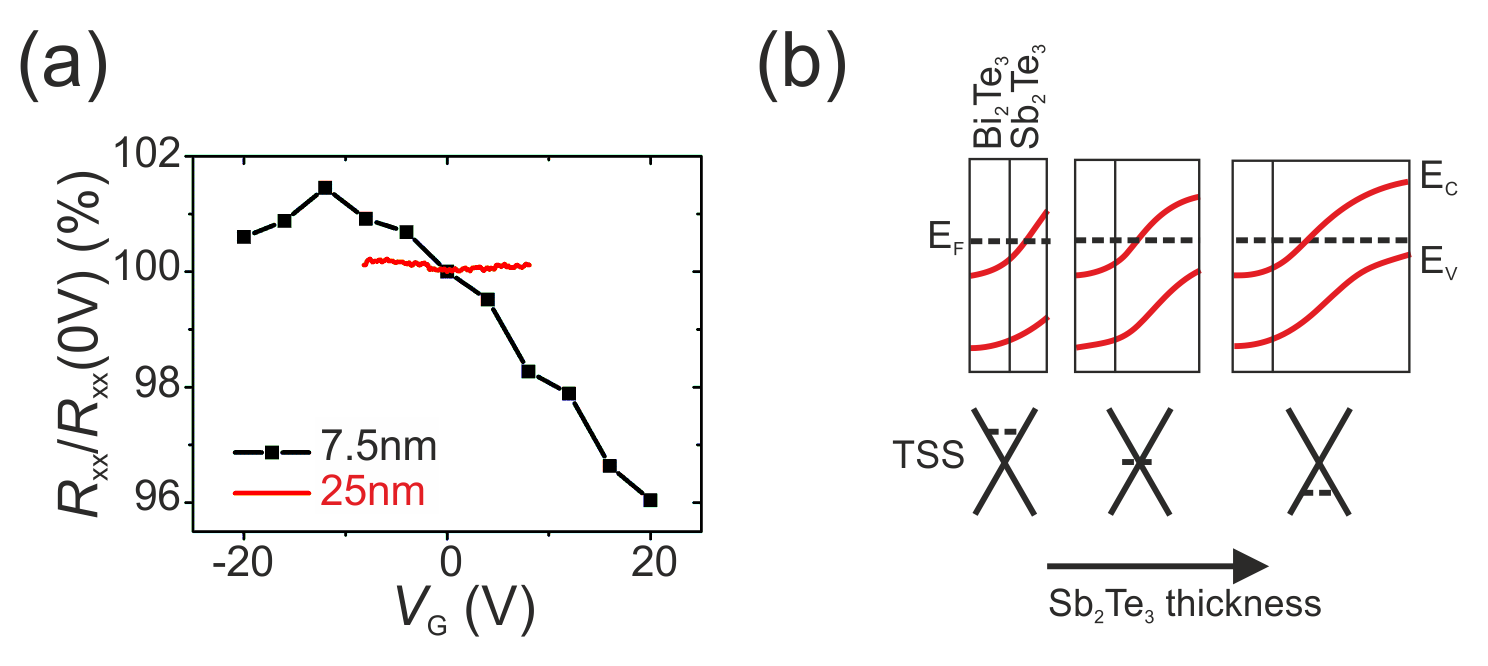} 
	\caption{(a) Gate voltage dependence of the resistivity for BST7 (black) and BST25 (red). (b) Schematic of the change of band structure as $t_\mathrm{SbTe}$ is increased.}
	\label{fig4}
\end{figure}

To test this conclusion we measure samples with top-gate electrodes which enable the tuning of the Fermi level $E_\mathrm{F}$ via a gate voltage $V_\mathrm{G}$. A variation of $E_\mathrm{F}$ should lead to perceptible changes of the transport properties of the TSS [see Fig.~\ref{fig4}(b)] while transport through the bulk should be less affected due to screening. As can be seen in Fig.~\ref{fig4}(a) this is indeed the case, with the resistance of the thin, TSS-dominated sample much more dependent on $V_\mathrm{G}$ than the thick, bulk-dominated sample. The resistance of the thin sample is maximized when $V_\mathrm{G}=-12\,V$, likely corresponding to the alignment of $E_\mathrm{F}$ with the Dirac point. Thus, broadly speaking, despite the basic nature of the model, it captures the essential physics and provides a consistent explanation of the dependence of the longitudinal and Hall transport components. Furthermore, the results of our calculation are clearly consistent with the observation of `no' Hall slope in BST7.

\subsection{Semi-classical theory}

Although our simplistic model offers useful physical insights, for a more microscopic understanding it is desirable that one is not dependent on \textit{ad-hoc} assumptions and/or a large number of experimental parameters. In the following we present a semi-classical theory for calculating magneto-conductivity tensors of surface and bulk charge carriers in a topological $p$-$n$ junction using zeroth and first-order Boltzmann moment equations~\cite{Huang:2004}. Assuming the $p$-$n$ interface to be in the $x-y$ plane, then under a parallel external electric field ${\bf E}=(E_\mathrm{x},E_\mathrm{y},0)$ and a perpendicular magnetic field ${\bf B}=(0,0,B)$, the total current per length in a $p$-$n$ junction structure is given by
$\displaystyle{\int_{-L_\mathrm{A}}^{L_\mathrm{D}} dz\,\left[{\bf j}^\|_\mathrm{c}(z)+{\bf j}^\|_\mathrm{v}(z)\right]+{\bf j}_\mathrm{s}^{\pm}}$, where $L_\mathrm{D}$ and $L_\mathrm{A}$ are the thickness of the $p$ region (donors) and $n$ region (acceptors), respectively. Here ${\bf j}_\mathrm{i}$ indicate the current densities with $i = c$, $v$ or $s$ for conduction band, valence band and surface, respectively. The superscript $\|$ is included to emphasise that the current considered is parallel to the $p$-$n$ interface as is experimentally the case. The bulk current densities are given by

\begin{widetext}
\begin{equation}
{\bf j}_\mathrm{c,v}^\|(z)=\frac{2e\gamma_\mathrm{e,h}m_\mathrm{e,h}^\ast\tau_\mathrm{e,h}(z)}{\tau_\mathrm{p(e,h)}(z)}\,{\bf v}^\|_\mathrm{c,v}[u_\mathrm{c,v}(z)]
\left\{\left[\tensor{\mbox{\boldmath${\mu}$}}^\|_\mathrm{c,v}({\bf B},z)\cdot{\bf E}\right]\right\}\cdot{\bf v}^\|_\mathrm{c,v}[u_\mathrm{c,v}(z)]\,
{\cal D}_\mathrm{c,v}[u_\mathrm{c,v}(z)]\ ,
\label{eqn1}
\end{equation}
\end{widetext}

\noindent where $\gamma_\mathrm{e,h}=-1$ or $+1$ for electrons and holes, respectively, 
$m^*_\mathrm{e,h}$ are effective masses of electrons and holes, 
$\tau_\mathrm{e,h}(z)$ and $\tau_\mathrm{p(e,h)}(z)$ are bulk energy- and momentum relaxation times~\cite{Huang:2004}, 
the velocity ${\bf v}^\|_\mathrm{c,v}({\bf k})=-\gamma_\mathrm{e,h}\,\hbar{\bf k}_\|/m^\ast_\mathrm{e,h}$ (with ${\bf k}$ the wave vector and ${\bf k}_{\|}$ the in-plane wave vector),
$u_\mathrm{c,v}(z)=(\hbar k_\mathrm{F}^\mathrm{e,h})^2/2m^*_\mathrm{e,h}$ and $k_\mathrm{F}^\mathrm{e,h}$ are Fermi energies and wave vectors in the bulk, 
{\boldmath${\mu}^{\|}_\mathrm{c,v}$} are mobility tensors, and 
${\cal D}_\mathrm{c,v}[u_\mathrm{c,v}(z)]=(\sqrt{u_\mathrm{c,v}(z)}/4\pi^2)\,(2m_\mathrm{e,h}^\ast/\hbar^2)^{3/2}$ is the electron and hole density-of-states per spin.

Similarly, one obtains the surface current per length as	

\begin{equation}
{\bf j}^{\pm}_\mathrm{s}=\mp\frac{e\tau_\mathrm{s}\hbar k^\mathrm{s}_\mathrm{F}}{\tau_\mathrm{sp}v_\mathrm{F}}\,{\bf v}^{\pm}_\mathrm{s}(u_\mathrm{s})
\left\{\left[\tensor{\mbox{\boldmath${\mu}$}}^{\pm}_\mathrm{s}({\bf B})\cdot{\bf E}\right]\right\}\cdot{\bf v}^{\pm}_\mathrm{s}(u_\mathrm{s})\,
\rho_\mathrm{s}(u_\mathrm{s})\ ,
\label{eqn2}
\end{equation}

\noindent where the $\pm$ denote when the Fermi level lies above and below the Dirac point, respectively, 
$\tau_\mathrm{s}$ and $\tau_\mathrm{sp}$ are surface energy- and momentum relaxation times, 
$k^\mathrm{s}_\mathrm{F}=\sqrt{4\pi n_\mathrm{s}}$ where $n_\mathrm{s}$ is the areal density of surface electrons,
$v_\mathrm{F}$ is the Fermi velocity of a Dirac cone, 
${\bf v}^{\pm}_\mathrm{s}({\bf k}_\|)=\pm({\bf k}_\|/k_\|)\,v_\mathrm{F}$,
$u_\mathrm{s}=\hbar v_\mathrm{F}k_\mathrm{F}^\mathrm{s}$ is the Fermi energy of a Dirac cone, and
$\rho_\mathrm{s}(u_\mathrm{s})=u_\mathrm{s}/(2\pi\hbar^2v_\mathrm{F}^2)$ is the surface density-of-states of a Dirac cone.

\noindent The bulk mobility tensors $\tensor{\mbox{\boldmath${\mu}$}}_\mathrm{c,v}({\bf B},z)$ are given by

\begin{equation}
\tensor{\mbox{\boldmath${\mu}$}}^\|_\mathrm{c,v}({\bf B},z)=\frac{\mu_\mathrm{0}(z)}{1+\mu^2_\mathrm{0}(z)B^2}\,
\left[\begin{array}{cc}
1 & \mu_\mathrm{0}(z)B\\
-\mu_\mathrm{0}(z)B & 1
\end{array}\right]\ ,
\label{eqn3}
\end{equation}
where $\mu_\mathrm{0}(z)=e\gamma_\mathrm{e,h}\tau_\mathrm{p{(e,h)}}(z)/m_\mathrm{e,h}^\ast$. A derivation of the bulk mobility tensor can be found in Appendix D.
The bulk conductivity tensor is then calculated as

\begin{multline}
\tensor{\mbox{\boldmath${\sigma}$}}^\|_\mathrm{c,v}({\bf B})=\\
e\gamma_\mathrm{e,h}\,\int_{-L_\mathrm{A}}^{L_\mathrm{D}} dz\,n_\mathrm{e,h}(z)\,\left[\frac{\tau_\mathrm{e,h}(z)}{\tau_\mathrm{p(e,h)}(z)}\right]\,\tensor{\mbox{\boldmath${\mu}$}}^\|_\mathrm{c,v}({\bf B},z)\ .
\label{eqn4}
\end{multline}
Likewise, the surface mobility tensor is

\begin{equation}
\tensor{\mbox{\boldmath${\mu}$}}^{\pm}_s({\bf B})=\mp\frac{\mu_1}{1+\mu^2_1B^2}
\left[\begin{array}{cc}
1 & \mp\mu_1B\\
\pm\mu_1B & 1\\
\end{array}\right]\ ,
\label{eqn5}
\end{equation}
where $\mu_1=4\epsilon^2_0\epsilon^2_\mathrm{r}\hbar v_\mathrm{F}^2/\sigma_\mathrm{i}e^3$, $\epsilon_\mathrm{r}$ is the host dielectric constant, 
and $\sigma_\mathrm{i}$ is the surface density of impurities. This corresponds to a surface conductivity tensor given by

\begin{equation}
\tensor{\mbox{\boldmath${\sigma}$}}^{\pm}_\mathrm{s}({\bf B})=e\sigma_\mathrm{s}\left(\frac{\tau_\mathrm{s}}{\tau_\mathrm{sp}}\right)\tensor{\mbox{\boldmath${\mu}$}}^{\pm}_\mathrm{s}({\bf B})\ .
\label{eqn6}
\end{equation}
Therefore, the total conductivity tensor $\tensor{\mbox{\boldmath${\sigma}$}}_\mathrm{tot}({\bf B})=\tensor{\mbox{\boldmath${\sigma}$}}^\|_\mathrm{c}({\bf B})+\tensor{\mbox{\boldmath${\sigma}$}}^\|_\mathrm{v}({\bf B})
+\tensor{\mbox{\boldmath${\sigma}$}}^{\pm}_\mathrm{s}({\bf B})$ is obtained as

\begin{widetext}
\begin{eqnarray}
\tensor{\mbox{\boldmath${\sigma}$}}_\mathrm{tot}({\bf B})=e\,\tensor{\mbox{\boldmath${\mu}$}}^\|_\mathrm{v}({\bf B})N_\mathrm{A}A_\mathrm{h}\left[(L_\mathrm{A}-W_\mathrm{p})+\int^{W_\mathrm{p}}_{0} dz\,\exp\left(-\frac{\beta e\bar{\mu}_\mathrm{h}N_\mathrm{A}}{2\epsilon_0\epsilon_\mathrm{r}D_\mathrm{h}}\,z^2\right)\right]
-e\,\tensor{\mbox{\boldmath${\mu}$}}^\|_\mathrm{c}({\bf B})N_\mathrm{D}A_\mathrm{e}\nonumber\\
\times\left[(L_\mathrm{D}-W_\mathrm{n})+\int^{W_\mathrm{n}}_{0} dz\,\exp\left(-\frac{\beta e\bar{\mu}_\mathrm{e}N_\mathrm{D}}{2\epsilon_0\epsilon_\mathrm{r}D_\mathrm{e}}\,z^2\right)\right]+e\,\tensor{\mbox{\boldmath${\mu}$}}^{\pm}_\mathrm{s}({\bf B})\,\left(\frac{\alpha_0^2}{4\pi\hbar^2v_\mathrm{F}^2}\right)\left(L_\mathrm{A}-L_\mathrm{0}\right)^2A_\mathrm{s}\ ,
\label{eqn7}
\end{eqnarray}
\end{widetext}

\noindent where $\alpha_0$ and $L_0$ are constants to be determined experimentally, $N_\mathrm{D,A}$ are doping concentrations, 
$W_\mathrm{n}$ and $W_\mathrm{p}$ are the thicknesses of the depletion zones for donors and acceptors in a $p$-$n$ junction,
$\bar{\mu}_\mathrm{e,h}$ are $\mu_0(z)$ evaluated at $n_\mathrm{e,h}(z)=N_\mathrm{D,A}$, $D_\mathrm{e,h}$ are diffusion coefficients, 
$\beta=4/3$ ($\beta=7/3$) for longitudinal (Hall) conductivity. In addition,
the averaged mobilities $\tensor{\mbox{\boldmath${\mu}$}}^\|_\mathrm{c,v}({\bf B})$ are defined by 
their values of $\tau_\mathrm{p(e,h)}(z)$ at $n_\mathrm{e,h}(z)=N_\mathrm{D,A}$, and three coefficients are $A_\mathrm{s}=\tau_\mathrm{s}/\tau_\mathrm{sp}\approx 3/4$, 

\begin{eqnarray}
A_\mathrm{e,h}&=&\left.\frac{\tau_\mathrm{e,h}(z)}{\tau_\mathrm{p(e,h)}(z)}\right|_{n_\mathrm{e,h}(z)=N_\mathrm{D,A}}\\
&=&\frac{1}{6}\left(\frac{Q_\mathrm{c}}{k_\mathrm{F}^\mathrm{e,h}}\right)^2\left[2\ln\left(\frac{2k_\mathrm{F}^\mathrm{e,h}}{Q_\mathrm{c}}\right)-1\right]\nonumber\\
&=&\frac{Q^2_\mathrm{c}}{6(3\pi^2N_\mathrm{D,A})^{2/3}}\left\{2\ln\left[\frac{2(3\pi^2N_\mathrm{D,A})^{1/3}}{Q_\mathrm{c}}\right]-1\right\}\nonumber ,
\label{eqn8}
\end{eqnarray}
where $1/Q_\mathrm{c}$ is the Thomas-Fermi screening length. More details on the derivation of the conductivity tensors can be found in Appendix E.

From Eq.~(\ref{eqn7}) one can see that there exists a critical value of $L_\mathrm{A} = L^*$  at which the total Hall conductivity becomes zero, which is determined from the following quadratic equation

\begin{widetext}
\begin{eqnarray}
\frac{\bar{\mu}^2_\mathrm{h}N_\mathrm{A}A_\mathrm{h}}{1+\bar{\mu}_\mathrm{h}^2B^2}\left\{(L^*-W_\mathrm{p})+\int^{W_\mathrm{p}}_{0} dz\,\exp\left[-\left(\frac{7e\bar{\mu}_\mathrm{h}N_\mathrm{A}}{6\epsilon_0\epsilon_\mathrm{r}D_\mathrm{h}}\right)z^2\right]\right\}
-\frac{\bar{\mu}^2_\mathrm{e}N_\mathrm{D}A_\mathrm{e}}{1+\bar{\mu}^2_\mathrm{e}B^2}\left\{(L_\mathrm{D}-W_\mathrm{n})\right.\nonumber\\
\left.+\int^{W_\mathrm{n}}_{0} dz\,\exp\left[-\left(\frac{7e\bar{\mu}_\mathrm{e}N_\mathrm{D}}{6\epsilon_0\epsilon_\mathrm{r}D_\mathrm{e}}\right)z^2\right]\right\}
\pm\frac{\mu_1^2}{1+\mu_1^2B^2}\left(\frac{\alpha_0^2}{4\pi\hbar^2v_\mathrm{F}^2}\right)\left(L^*-L_0\right)^2A_\mathrm{s}=0\ ,
\label{eqn9}
\end{eqnarray}
\end{widetext}
where the sign $+$ ($-$) corresponds to $L_\mathrm{A}>L_0$ ($L_\mathrm{A}<L_0$) for the contribution of the lower (upper) Dirac cone.

We note that in arriving at the above equations we have not considered scattering between the TSS and bulk layers. Including these will modify energy-relaxation times for both bulk and surface states, although no analytical expression for these can be obtained even at low $T$. We leave a numerical evaluation of the problem for a later paper. For the purposes of this paper, we stress that the inclusion of this coupling only serves to modify the three coefficients $A_\mathrm{e}$, $A_\mathrm{h}$, and $A_\mathrm{s}$, and thus the obtained result is qualitatively unchanged. Importantly, the physical content of Eq.~(\ref{eqn9}) is essentially identical to that in Eq.~(\ref{Hall}), but arrived at in a more rigorous fashion. This provides a very useful microscopic grounding to Eq.~(\ref{Hall}) while also providing additional confidence to the physical insights drawn from the simple three-channel model.

\section{Conclusion}

In conclusion, we have reported low-$T$ magnetotransport measurements on vertical topological $p$-$n$ junctions and understood the data within a three-channel model for the Hall resistance. It provides useful insights into the complex interplay of the bulk and TSS in the multilayered TI, explains the sign change of $R_\mathrm{H}$ with varying $t_\mathrm{SbTe}$, and delivers values for the mobility of the TSS of 281\,$\mathrm{cm^2V^{-1}s^{-1}}$. We then develop a Boltzmann transport theory which provides a clear microscopic foundation for our model. Our work paves the way for the study of other complex TI heterostructures~\cite{Narayan:2016,Nguyen:2016, Belopolski:2017}, where bulk states and TSS of different carrier types coexist. In future, our method can be applied to improved topological $p$-$n$ junctions in which a top and bottom TSS can form novel Dirac fermion excitonic states.

\begin{acknowledgments}
D.B., D.R. and V.N. acknowledge funding from the Leverhulme Trust, UK, D.B., R.M., D.R., and V.N. acknowledge funding from EPSRC (UK). DH would like to thank the support from the Air Force Office of Scientific Research (AFOSR). G.M., M.L., J.K. and D.G. acknowledge financial support from the DFG-funded priority programme SPP1666. 

Supporting data for this paper is available at the DSpace@Cambridge data repository (https://doi.org/10.17863/CAM.13094).
\end{acknowledgments}


\appendix

\section{ERROR ESTIMATES FOR $\alpha$}

Figure~\ref{fig2}(a) compares the results when 1) $\alpha$ and $l_{\phi}$ were both fitting variables (red line) or 2) when $l_{\phi}$ alone was used as a fitting variable and $\alpha$ was kept constant. We find that the fit for $\alpha = 1$ (blue dashed-dotted line) is of a significantly poorer quality, indicating clearly that the data are consistent with the existence of one WAL mode. These errors become significantly larger as $T$ is increased (here not shown) and thus one must not over interpret the apparent increase in $\alpha$ with $T$ in Fig.~\ref{fig2}(d).

\section{TSS ELECTRON DENSITY}

The density of states in the dirac cone \cite{Kittel:1986} is given by  

\begin{equation}
g(k)dk/\left(\frac{2 \pi}{L}\right)^2=2\pi k  dk/\left(\frac{2 \pi}{L}\right)^2=\frac{k dk}{2 \pi/L^2}.
\end{equation}

The relation between the binding energy $E_\mathrm{B}$, i.e. the difference between the Fermi energy and the Dirac point, and the Fermi wave vector $k_\mathrm{F}$ is

\begin{equation}
E_\mathrm{B}=\beta k_\mathrm{F}=\hbar v_\mathrm{F} k_\mathrm{F}
\end{equation} 

and can be retrieved from ARPES measurements in Ref.~\onlinecite{Eschbach:2015}, carried out using samples from the same growth process and identical material parameters. For $E_\mathrm{B}=215$\,meV, $k_\mathrm{F}\approx 0.1 \mathrm{\AA}$ [see Fig. 4(h) in Ref.~\onlinecite{Eschbach:2015}], thus $\beta=\frac{E_\mathrm{B}}{k_\mathrm{F}}=3.44\times10^{-29}$J m. From $\beta$, a Fermi velocity of $3.26\times10^5\,\mathrm{m/s}$ can be derived.  

The electron density of the TSS is  

\begin{equation}
n_\mathrm{t}=k_\mathrm{F}^2/4\pi=\frac{E_\mathrm{B}^2}{4\pi \beta^2 }.
\label{eqnb3}
\end{equation}

Furthermore, the relation between $E_\mathrm{B}$ and the $\mathrm{Sb_2Te_3}$-thickness is linear $(dE_\mathrm{B}/dt_\mathrm{SbTe}=1.62\times 10^{-12}$\,J/m, see Fig.~\ref{SMfig1}) and 

\begin{equation}
\label{nt}
n_\mathrm{t}=\frac{(dE_\mathrm{B}/dt_\mathrm{SbTe}\cdot t_\mathrm{SbTe})^2}{4\pi \beta^2}.
\end{equation}

\section{DERIVATION OF $R_\mathrm{H}$ AND $n_\mathrm{eff}$}

The force acting on charges in the TSS (index t), bulk-$\mathrm{Sb_2Te_3}$ (p) and bulk-$\mathrm{Bi_2Te_3}$ (n) originate from an electric field $\vec{E}$ in y-direction and a magnetic field $\vec{B}$ in z-direction:

\begin{equation}
\begin{split}
-F_\mathrm{ny} &=e E_\mathrm{y}+ev_\mathrm{nx}B_\mathrm{z}\\
-F_\mathrm{ty} &=e E_\mathrm{y}+ev_\mathrm{tx}B_\mathrm{z}\\
F_\mathrm{py} &=e E_\mathrm{y}-ev_\mathrm{px}B_\mathrm{z} 
\end{split}
\end{equation}

Using $v=\frac{\mu}{e}F$ with $\mu$ the mobility, we obtain

\begin{equation}
\begin{split}
\frac{v_\mathrm{ny}}{\mu_\mathrm{n}} &=E_\mathrm{y}+\mu_\mathrm{n}E_\mathrm{x}B_\mathrm{z}\\
\frac{v_\mathrm{ty}}{\mu_\mathrm{t}} &=E_\mathrm{y}+\mu_\mathrm{t}E_\mathrm{x}B_\mathrm{z}\\
\frac{v_\mathrm{py}}{\mu_\mathrm{p}} &=E_\mathrm{y}-\mu_\mathrm{p}E_\mathrm{x}B_\mathrm{z}
\end{split}
\end{equation}

Furthermore, no charge current is flowing in y-direction

\begin{equation}
\begin{split}
J_\mathrm{y} &=J_\mathrm{n}+J_\mathrm{t}+J_\mathrm{p} \\
&=en_\mathrm{n}v_\mathrm{ny}+en_\mathrm{t}v_\mathrm{ty}+e\mathrm{n}_\mathrm{p}\mathrm{v}_\mathrm{py}  =0\\
      \implies n_\mathrm{n}v_\mathrm{ny}&=-(n_\mathrm{t}v_\mathrm{ty}+n_\mathrm{p}v_\mathrm{py})  \\
\end{split}
\end{equation}

Inserting the velocities in the previous equation gives

\begin{equation}
\begin{split}
&n_\mathrm{n}\mu_\mathrm{n} (E_\mathrm{y}+\mu_\mathrm{n} E_\mathrm{x} B_\mathrm{z}) \\
&=-(n_\mathrm{t} \mu_\mathrm{t} (E_\mathrm{y}+ \mu_\mathrm{t}E_\mathrm{x}B_\mathrm{z})+n_\mathrm{p}\mu_\mathrm{p} (E_\mathrm{y}-\mu_\mathrm{p}E_\mathrm{x}B_\mathrm{z}))\\
\implies &E_\mathrm{y} (n_\mathrm{n}\mu_\mathrm{n}+n_\mathrm{t}\mu_\mathrm{t}+n_\mathrm{p}\mu_\mathrm{p})\\
&=B_\mathrm{z}E_\mathrm{x}(-n_\mathrm{n}\mu_\mathrm{n}^2-n_\mathrm{t}\mu_\mathrm{t}^2+n_\mathrm{p}\mu_\mathrm{p}^2)
\end{split}
\end{equation}

The charge current in x-direction is

\begin{equation}
\begin{split}
J_\mathrm{x}&=en_\mathrm{n}v_\mathrm{nx}+en_\mathrm{t}v_\mathrm{tx}+en_\mathrm{p}v_\mathrm{px}\\
&=(n_\mathrm{n}\mu_\mathrm{n}+n_\mathrm{t}\mu_\mathrm{t}+n_\mathrm{p}\mu_\mathrm{p})eE_\mathrm{x}
\end{split}
\end{equation}

$E_\mathrm{x}$ can now be replaced, resulting in

\begin{equation}
\begin{split}
&eE_\mathrm{y}(n_\mathrm{n}\mu_\mathrm{n}+n_\mathrm{t}\mu_\mathrm{t}+n_\mathrm{p}\mu_\mathrm{p})^2\\
&=B_\mathrm{z}J_\mathrm{x}(-n_\mathrm{n}\mu_\mathrm{n}^2-n_\mathrm{t}\mu_\mathrm{t}^2+n_\mathrm{p}\mu_\mathrm{p}^2)\\
&\implies R_\mathrm{H}=\frac{B_\mathrm{z}J_\mathrm{x}}{E_\mathrm{y}}=\frac{-n_\mathrm{n}\mu_\mathrm{n}^2-n_\mathrm{t}\mu_\mathrm{t}^2+n_\mathrm{p}\mu_\mathrm{p}^2}{e(n_\mathrm{n}\mu_\mathrm{n}+n_\mathrm{t}\mu_\mathrm{t}+n_\mathrm{p}\mu_\mathrm{p})^2}
\end{split}
\end{equation}

\begin{figure}
	\centering
	\includegraphics[width=\columnwidth]{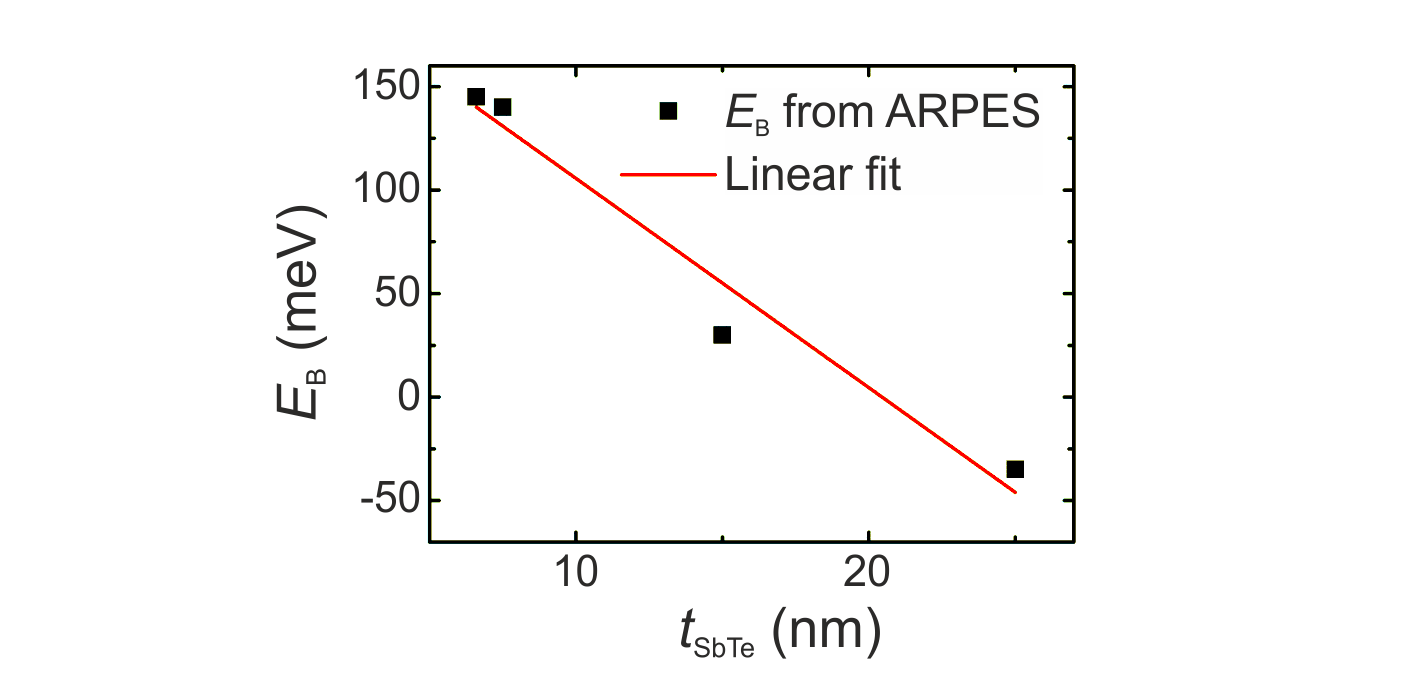}
	\caption{Relation between $E_\mathrm{B}$ and $t_\mathrm{SbTe}$ (from Ref.~\onlinecite{Eschbach:2015})}
	\label{SMfig1}
\end{figure}

Both $n_\mathrm{p}$ and $n_\mathrm{t}$ are depending on the thickness of the $\mathrm{Sb_2Te_3}$-thickness, $t_\mathrm{SbTe}$, with 

\begin{equation}
\begin{split}
n_\mathrm{p}&=n_\mathrm{SbTe}\cdot t_\mathrm{SbTe}\\
n_\mathrm{t}(t_\mathrm{SbTe})&=\frac{(dE_\mathrm{B}/dt_\mathrm{SbTe}\cdot (t_\mathrm{SbTe}-t_\mathrm{0}))^2}{4\pi \beta^2}
\end{split}
\end{equation}

\noindent where $dE_\mathrm{B}/dt_\mathrm{SbTe}$ can be gained from Fig. \ref{SMfig1}. 

Thus $R_\mathrm{H}(t_\mathrm{SbTe})$ is a function of the $\mathrm{Sb_2Te_3}$-thickness of the form

\begin{equation}
\begin{split}
R_\mathrm{H}(t_\mathrm{SbTe})=\frac{-n_\mathrm{n}(t_\mathrm{SbTe})\mu_\mathrm{n}^2\pm n_\mathrm{t}(t_\mathrm{SbTe})\mu_\mathrm{t}^2+n_\mathrm{p}\mu_\mathrm{p}^2}{e(n_\mathrm{n}(t_\mathrm{SbTe})\mu_\mathrm{n}+n_\mathrm{t}(t_\mathrm{SbTe})\mu_\mathrm{t}+n_\mathrm{p}\mu_\mathrm{p})^2}\\
=\frac{-n_\mathrm{SbTe}t_\mathrm{SbTe}\mu_\mathrm{n}^2\pm \frac{(dE_\mathrm{B}/dt_\mathrm{SbTe}\cdot (t_\mathrm{SbTe}-t_\mathrm{0}))^2}{4\pi \beta^2}\mu_\mathrm{t}^2+n_\mathrm{p}\mu_\mathrm{p}^2}{e(n_\mathrm{SbTe}t_\mathrm{SbTe}\mu_\mathrm{n}+\frac{(dE_\mathrm{B}/dt_\mathrm{SbTe}\cdot (t_\mathrm{SbTe}-t_\mathrm{0}))^2}{4\pi \beta^2}\mu_\mathrm{t}+n_\mathrm{p}\mu_\mathrm{p})^2}
\end{split}
\end{equation}

\noindent where the `+' sign has to be used when $t_\mathrm{SbTe}>20$\,nm and the `-' sign for $t_\mathrm{SbTe}<20$\,nm.

Because of the entity $R_\mathrm{H}=-1/(e \cdot n_\mathrm{eff})$, the ``effective'' two-dimensional charge density is given by

\begin{equation}
\begin{split}
n_\mathrm{eff}&=-\frac{[n_\mathrm{n}(t_\mathrm{SbTe})\mu_\mathrm{n}+n_\mathrm{t}(t_\mathrm{SbTe})\mu_\mathrm{t}+n_\mathrm{p}\mu_\mathrm{p}]^2}{-n_\mathrm{n}(t_\mathrm{SbTe})\mu_\mathrm{n}^2\pm n_\mathrm{t}(t_\mathrm{SbTe})\mu_\mathrm{t}^2+n_\mathrm{p}\mu_\mathrm{p}^2}
\end{split}
\end{equation}

\section{BULK AND SURFACE MOBILITY TENSORS}

By using the force-balance equation\,\cite{ref-1,ref-2,Huang:2004} for bulk electrons

\begin{multline}
\frac{\partial\mbox{\boldmath$v$}_\mathrm{d}(t\vert z)}{\partial t}
=-\tensor{\mbox{\boldmath$\tau$}}_\mathrm{pe}^{-1}(z)\cdot\mbox{\boldmath$v$}_\mathrm{d}(t\vert z)\\
-e\tensor{\mbox{\boldmath${\cal M}$}}_\mathrm{c}^{-1}(z)\cdot\left[{\bf E}(t)+\mbox{\boldmath$v$}_\mathrm{d}(t\vert z)\times{\bf B}(t)\right]=0\ ,
\label{app-1}
\end{multline}
as well as the diagonal approximation for the inverse momentum-relaxation-time tensor $\tensor{\mbox{\boldmath${\tau}$}}_\mathrm{pe}^{-1}\approx(1/\tau_\mathrm{j})\,\delta_\mathrm{ij}$,
we get the following group of linear inhomogeneous equations for $\mbox{\boldmath$v$}_\mathrm{d}=\{v_1,\,v_2,\,v_3\}$

\begin{multline}
\left[1+q\tau_1\left(r_{12}B_3-r_{13}B_2\right)\right]v_1+q\tau_1\left(r_{13}B_1-r_{11}B_3\right)v_2\\
+q\tau_1\left(r_{11}B_2-r_{12}B_1\right)v_3
=q\tau_1\left(r_{11}E_1+r_{12}E_2+r_{13}E_3\right)\ ,\\
%
q\tau_2\left(r_{22}B_3-r_{23}B_2\right)v_1+\left[1+q\tau_2\left(r_{23}B_1-r_{21}B_3\right)\right]v_2\\
+q\tau_2\left(r_{21}B_2-r_{22}B_1\right)v_3
=q\tau_2\left(r_{21}E_1+r_{22}E_2+r_{23}E_3\right)\ ,\\
%
q\tau_3\left(r_{32}B_3-r_{33}B_2\right)v_1+q\tau_3\left(r_{33}B_1-r_{31}B_3\right)v_2+\\
\left[1+q\tau_3\left(r_{31}B_2-r_{32}B_1\right)\right]v_3
=q\tau_3\left(r_{31}E_1+r_{32}E_2+r_{33}E_3\right)\ ,
\label{app-4}
\end{multline}
where the statistically-averaged inverse effective-mass tensor for the conduction band is

\begin{multline}
\left[\tensor{\mbox{\boldmath${\cal M}$}}_\mathrm{c}^{-1}(z)\right]_\mathrm{ij}\equiv\{r_\mathrm{ij}\}\equiv\\
\frac{2}{n_\mathrm{e}(z){\cal V}}\,\sum_{{\bf k}}\,\left[\frac{1}{\hbar^2}\,\frac{\partial^2\varepsilon_\mathrm{c}({\bf k})}{\partial k_\mathrm{i}\partial k_\mathrm{j}}\right]f_0[\varepsilon_\mathrm{c}({\bf k}),T;\,u_\mathrm{c}(z)]\ ,
\label{app-5}
\end{multline}
$i,\,j=x,\,y,\,z$, ${\bf B}=\{B_1,\,B_2,\,B_3\}$, ${\bf E}=\{E_1,\,E_2,\,E_3\}$, and $q=-e$.
By defining the coefficient matrix $\tensor{\mbox{\boldmath${\cal C}$}}$ for the above linear equations, i.e.,

\begin{widetext}
\begin{equation}
\tensor{\mbox{\boldmath${\cal C}$}}=
\left[\begin{array}{ccc}
1+q\tau_1(r_{12}B_3-r_{13}B_2) & q\tau_1(r_{13}B_1-r_{11}B_3) & q\tau_1(r_{11}B_2-r_{12}B_1)\\
q\tau_2(r_{22}B_3-r_{23}B_2) & 1+q\tau_2(r_{23}B_1-r_{21}B_3) & q\tau_2(r_{21}B_2-r_{22}B_1)\\
q\tau_3(r_{32}B_3-r_{33}B_2) & q\tau_3(r_{33}B_1-r_{31}B_3) & 1+q\tau_3(r_{31}B_2-r_{32}B_1)
\end{array}\right]\ ,
\label{app-6}
\end{equation}
\end{widetext}
as well as the source vector ${\bf s}$, given by

\begin{equation}
{\bf s}=\left[\begin{array}{c}
q\tau_1(r_{11}E_1+r_{12}E_2+r_{13}E_3)\\
q\tau_2(r_{21}E_1+r_{22}E_2+r_{23}E_3)\\
q\tau_3(r_{31}E_1+r_{32}E_2+r_{33}E_3)
\end{array}\right]\ ,
\label{app-7}
\end{equation}
we can reduce the linear equations to a matrix equation $\tensor{\mbox{\boldmath${\cal C}$}}\cdot{\bf v}_\mathrm{d}={\bf s}$ with a formal solution
$\mbox{\boldmath$v$}_\mathrm{d}=\tensor{\mbox{\boldmath${\cal C}$}}^{-1}\cdot{\bf s}$. Explicitly, we find the solution $\mbox{\boldmath$v$}_\mathrm{d}=\{v_1,\,v_2,\,v_3\}$
for $j=1,\,2,\,3$ as

\begin{equation}
v_\mathrm{j}=\frac{\mathrm{Det}\{\tensor{\mbox{\boldmath${\Delta}$}}_\mathrm{j}\}}{\mathrm{Det}\{\tensor{\mbox{\boldmath${\cal C}$}}\}}\ ,
\label{app-8}
\end{equation}
where $\mathrm{Det}\{\cdots\}$ means taking the determinant,

\footnotesize
\begin{widetext}
\begin{equation*}
\tensor{\mbox{\boldmath${\Delta}$}}_1=
\left[\begin{array}{ccc}
q\tau_1(r_{11}E_1+r_{12}E_2+r_{13}E_3) & q\tau_1(r_{13}B_1-r_{11}B_3) & q\tau_1(r_{11}B_2-r_{12}B_1)\\
q\tau_2(r_{21}E_1+r_{22}E_2+r_{23}E_3) & 1+q\tau_2(r_{23}B_1-r_{21}B_3) & q\tau_2(r_{21}B_2-r_{22}B_1)\\
q\tau_3(r_{31}E_1+r_{32}E_2+r_{33}E_3) & q\tau_3(r_{33}B_1-r_{31}B_3) & 1+q\tau_3(r_{31}B_2-r_{32}B_1)
\end{array}\right]\ ,
\label{app-9}
\end{equation*}

\begin{equation}
\tensor{\mbox{\boldmath${\Delta}$}}_2=
\left[\begin{array}{ccc}
1+q\tau_1(r_{12}B_3-r_{13}B_2) & q\tau_1(r_{11}E_1+r_{12}E_2+r_{13}E_3) & q\tau_1(r_{11}B_2-r_{12}B_1)\\
q\tau_2(r_{22}B_3-r_{23}B_2) & q\tau_2(r_{21}E_1+r_{22}E_2+r_{23}E_3) & q\tau_2(r_{21}B_2-r_{22}B_1)\\
q\tau_3(r_{32}B_3-r_{33}B_2) & q\tau_3(r_{31}E_1+r_{32}E_2+r_{33}E_3) & 1+q\tau_3(r_{31}B_2-r_{32}B_1)
\end{array}\right]\ ,
\label{app-10}
\end{equation}

\begin{equation*}
\tensor{\mbox{\boldmath${\Delta}$}}_3=
\left[\begin{array}{ccc}
1+q\tau_1(r_{12}B_3-r_{13}B_2) & q\tau_1(r_{13}B_1-r_{11}B_3) & q\tau_1(r_{11}E_1+r_{12}E_2+r_{13}E_3)\\
q\tau_2(r_{22}B_3-r_{23}B_2) & 1+q\tau_2(r_{23}B_1-r_{21}B_3) & q\tau_2(r_{21}E_1+r_{22}E_2+r_{23}E_3)\\
q\tau_3(r_{32}B_3-r_{33}B_2) & q\tau_3(r_{33}B_1-r_{31}B_3) & q\tau_3(r_{31}E_1+r_{32}E_2+r_{33}E_3)
\end{array}\right]\ .
\label{app-11}
\end{equation*}
\end{widetext}
\medskip

\normalsize
By assuming $r_\mathrm{ij}=0$ for $i\neq j$, $r_\mathrm{jj}=1/m_\mathrm{j}^\ast$ and introducing the notation $\mu_\mathrm{j}=q\tau_\mathrm{j}/m_\mathrm{j}^\ast$, we find

\begin{equation}
\begin{split}
\tensor{\mbox{\boldmath${\cal C}$}}=&
\left[\begin{array}{ccc}
1 & -\mu_1B_3 & \mu_1B_2\\
\mu_2B_3 & 1 & -\mu_2B_1\\
-\mu_3B_2 & \mu_3B_1 & 1
\end{array}\right]\ ,\\
%
\tensor{\mbox{\boldmath${\Delta}$}}_1=&
\left[\begin{array}{ccc}
\mu_1E_1 & -\mu_1B_3 & \mu_1B_2\\
\mu_2E_2 & 1 & -\mu_2B_1\\
\mu_3E_3 & \mu_3B_1 & 1
\end{array}\right]\ ,\\
%
\tensor{\mbox{\boldmath${\Delta}$}}_2=&
\left[\begin{array}{ccc}
1 & \mu_1E_1 & \mu_1B_2\\
\mu_2B_3 & \mu_2E_2 & -\mu_2B_1\\
-\mu_3B_2 & \mu_3E_3 & 1
\end{array}\right]\ ,\\
%
\tensor{\mbox{\boldmath${\Delta}$}}_3=&
\left[\begin{array}{ccc}
1 & -\mu_1B_3 & \mu_1E_1\\
\mu_2B_3 & 1 & \mu_2E_2\\
-\mu_3B_2 & \mu_3B_1 & \mu_3E_3
\end{array}\right]\ ,\\
\label{app-15}
\end{split}
\end{equation}
and

\begin{equation}
\begin{split}
\mathrm{Det}\{\tensor{\mbox{\boldmath${\cal C}$}}\}=&1+(B_1^2\mu_2\mu_3+B_2^2\mu_3\mu_1+B_3^2\mu_1\mu_2)\ ,\\
\mathrm{Det}\{\tensor{\mbox{\boldmath${\Delta}$}}_1\}=&
\mu_1E_1+\mu_1(B_3E_2\mu_2-B_2E_3\mu_3)\\
&+\mu_1\mu_2\mu_3B_1({\bf E}\cdot{\bf B})\ ,\\
\mathrm{Det}\{\tensor{\mbox{\boldmath${\Delta}$}}_2\}=&
\mu_2E_2+\mu_2(B_1E_3\mu_3-B_3E_1\mu_1)\\
&+\mu_1\mu_2\mu_3B_2({\bf E}\cdot{\bf B})\ ,\\
\mathrm{Det}\{\tensor{\mbox{\boldmath${\Delta}$}}_3\}=&\mu_3E_3+\mu_3(B_2E_1\mu_1-B_1E_2\mu_2)\\
&+\mu_1\mu_2\mu_3B_3({\bf E}\cdot{\bf B})\ .\\
\end{split}
\end{equation}
\medskip

If we further assume $m_1^\ast=m_2^\ast=m_3^\ast=m_e^\ast$ and $\tau_1=\tau_2=\tau_3=\tau_\mathrm{pe}$, we obtain
$\mathrm{Det}\{\tensor{\mbox{\boldmath${\cal C}$}}\}=1+\mu^2_0B^2$,
$\mathrm{Det}\{\tensor{\mbox{\boldmath${\Delta}$}}_1\}=-\mu_0E_1+\mu_0^2(B_3E_2-B_2E_3)-\mu_0^3B_1({\bf E}\cdot{\bf B})$,
$\mathrm{Det}\{\tensor{\mbox{\boldmath${\Delta}$}}_2\}=-\mu_0E_2+\mu_0^2(B_1E_3-B_3E_1)-\mu_0^3B_2({\bf E}\cdot{\bf B})$, and
$\mathrm{Det}\{\tensor{\mbox{\boldmath${\Delta}$}}_3\}=-\mu_0E_3+\mu_0^2(B_2E_1-B_1E_2)-\mu_0^3B_3({\bf E}\cdot{\bf B})$,
where $\mu_\mathrm{0}=e\tau_\mathrm{pe}/m_\mathrm{e}^\ast$. As a result, the mobility tensor $\tensor{\mbox{\boldmath${\mu}$}}_\mathrm{c}({\bf B})$, which is defined through $\mbox{\boldmath$v$}_\mathrm{d}=\tensor{\mbox{\boldmath${\mu}$}}_\mathrm{c}({\bf B})\cdot{\bf E}$, can be written as

\begin{widetext}
\begin{equation}
\tensor{\mbox{\boldmath${\mu}$}}_\mathrm{c}({\bf B})=-\frac{\mu_0}{1+\mu^2_0B^2}\,
\left[\begin{array}{ccc}
1+\mu_0^2B_1^2 & -\mu_0B_3+\mu_0^2B_1B_2 & \mu_0B_2+\mu_0^2B_1B_3\\
\mu_0B_3+\mu_0^2B_2B_1 & 1+\mu_0^2B_2^2 & -\mu_0B_1+\mu_0^2B_2B_3\\
-\mu_0B_2+\mu_0^2B_3B_1 & \mu_0B_1+\mu_0^2B_3B_2 & 1+\mu_0^2B_3^2
\end{array}\right]\ ,
\label{app-16}
\end{equation}
\end{widetext}
where $B^2=B_1^2+B_2^2+B_3^2$. By taking ${\bf B}=\{0,\,0,\,B\}$, we find from Eq.\,(\ref{app-16}) that

\begin{equation}
\tensor{\mbox{\boldmath${\mu}$}}_\mathrm{c}({\bf B})=-\frac{\mu_0}{1+\mu^2_0B^2}\,
	\left[\begin{array}{ccc}
	1 & -\mu_0B & 0\\
	\mu_0B & 1 & 0\\
	0 & 0 & 1+\mu_0^2B^2
	\end{array}\right]\ .
\label{app-17}
\end{equation}
\medskip

For the surface case, $E_3=0$, $v_3=0$ and $\tensor{\mbox{\boldmath${\cal M}$}}_\mathrm{s}^{-1}$, $\tensor{\mbox{\boldmath${\tau}$}}_\mathrm{sp}^{-1}$ and $\tensor{\mbox{\boldmath${\mu}$}}_\mathrm{s}({\bf B})$
for the $E_\mathrm{s}^{-}({\bf k}_\|)$ (lower-cone) state all reduce to $2\times 2$ tensors.
This gives rise to

\begin{equation}
\tensor{\mbox{\boldmath${\mu}$}}_\mathrm{s}({\bf B})=\frac{\mu_1}{1+\mu^2_1B^2}\,
	\left[\begin{array}{cc}
	1 & \mu_1B\\
	-\mu_1B & 1\\
	\end{array}\right]\ ,
\label{app-18}
\end{equation}
where $\mu_1=e\tau_\mathrm{sp}v_\mathrm{F}/(\hbar k^\mathrm{s}_\mathrm{F})$, $k^\mathrm{s}_\mathrm{F}=\sqrt{4\pi \sigma_\mathrm{s}}$ and $\sigma_\mathrm{s}$ is the areal density of surface electrons.

\section{BULK AND SURFACE CONDUCTIVITY TENSORS}

Under a parallel external electric field ${\bf E}=(E_\mathrm{x},E_\mathrm{y},0)$ and a perpendicular magnetic field ${\bf B}=(0,0,B)$, 
the total parallel current per length in a $p$-$n$ junction structure is given by
$\displaystyle{\int_{-L_\mathrm{A}}^{L_\mathrm{D}} dz\,\left[{\bf j}^\|_\mathrm{c}(z)+{\bf j}^\|_\mathrm{v}(z)\right]+{\bf j}_\mathrm{s}^{\pm}}$, where $L_\mathrm{D}$ and $L_\mathrm{A}$ are the distribution ranges for donors and acceptors, respectively.
Here, by using the second-order Boltzmann moment equation\,\cite{ref-4}, the bulk current densities are found to be

\begin{widetext}
\begin{equation}
	{\bf j}_\mathrm{c,v}^\|(z)=
	\frac{2e\gamma_\mathrm{e,h}m_\mathrm{e,h}^\ast\tau_\mathrm{e,h}(z)}{\tau_\mathrm{p(e,h)}(z)}\,{\bf v}^\|_\mathrm{c,v}[u_\mathrm{c,v}(z)]\left\{\left[\tensor{\mbox{\boldmath${\mu}$}}^\|_\mathrm{c,v}({\bf B},z)\cdot{\bf E}\right]\right\}\cdot{\bf v}^\|_\mathrm{c,v}[u_\mathrm{c,v}(z)]\,
	{\cal D}_\mathrm{c,v}[u_\mathrm{c,v}(z)]\ ,
	\label{app-19}
\end{equation}
\end{widetext}
where ${\cal D}_\mathrm{c,v}[u_\mathrm{c,v}(z)]=(\sqrt{u_\mathrm{c,v}(z)}/4\pi^2)\,(2m_\mathrm{e,h}^\ast/\hbar^2)^{3/2}$ is the electron and hole density-of-states per spin, $u_\mathrm{c,v}(z)=(\hbar k_\mathrm{F}^\mathrm{e,h})^2/2m^*_\mathrm{e,h}$
and $k_\mathrm{F}^\mathrm{e,h}$ are Fermi energies and wave vectors in a bulk,
$m^*_\mathrm{e,h}$ are effective masses of electrons and holes,
$\tau_\mathrm{e,h}(z)$ and $\tau_\mathrm{p(e,h)}(z)$ are bulk energy- and momentum relaxation times,\,\cite{ref-1,ref-2,Huang:2004}
${\bf v}^\|_\mathrm{c,v}({\bf k})=-\gamma_\mathrm{e,h}\,\hbar{\bf k}_\|/m^\ast_\mathrm{e,h}$,
and $\gamma_\mathrm{e,h}=-1$ (electrons) and $+1$ (holes), respectively. Similarly, the surface current per length is\,\cite{ref-4}

\begin{equation}
	{\bf j}^{\pm}_\mathrm{s}=\mp\frac{e\tau_\mathrm{s}\hbar k^\mathrm{s}_\mathrm{F}}{\tau_\mathrm{sp}v_\mathrm{F}}\,{\bf v}^{\pm}_\mathrm{s}(u_\mathrm{s})
	\left\{\left[\tensor{\mbox{\boldmath${\mu}$}}^{\pm}_\mathrm{s}({\bf B})\cdot{\bf E}\right]\right\}\cdot{\bf v}^{\pm}_\mathrm{s}(u_\mathrm{s})\,
	\rho_\mathrm{s}(u_\mathrm{s})\ ,
	\label{app-20}
\end{equation}
where $\rho_\mathrm{s}(u_\mathrm{s})=u_\mathrm{s}/(2\pi\hbar^2v_\mathrm{F}^2)$ and $u_\mathrm{s}=\hbar v_\mathrm{F}k_\mathrm{F}^\mathrm{s}$ are the surface density-of-states and Fermi energy, $k^\mathrm{s}_\mathrm{F}=\sqrt{4\pi \sigma_\mathrm{s}}$,
$v_\mathrm{F}$ is the Fermi velocity of a Dirac cone, 
$\tau_\mathrm{s}$ and $\tau_\mathrm{sp}$ are surface energy- and momentum relaxation times,\,\cite{ref-1,ref-2,Huang:2004}
and ${\bf v}^{\pm}_\mathrm{s}({\bf k}_\|)=\pm({\bf k}_\|/k_\|)\,v_\mathrm{F}$.
\medskip

From Eq.\,(\ref{app-19}), we find the bulk conductivity tensor as

\begin{equation}
	\tensor{\mbox{\boldmath${\sigma}$}}^\|_\mathrm{c,v}({\bf B})
	=e\gamma_\mathrm{e,h}\,\int_{-L_\mathrm{A}}^{L_\mathrm{D}} dz\,n_\mathrm{e,h}(z)\,\left[\frac{\tau_\mathrm{e,h}(z)}{\tau_\mathrm{p(e,h)}(z)}\right]\,\tensor{\mbox{\boldmath${\mu}$}}^\|_\mathrm{c,v}({\bf B},z)\ .
	\label{app-21}
\end{equation}

On the other hand, from Eq.\,(\ref{app-20}) we get the surface conductivity tensor, given by

\begin{equation}
	\tensor{\mbox{\boldmath${\sigma}$}}^{\pm}_\mathrm{s}({\bf B})=e\sigma_\mathrm{s}\left(\frac{\tau_\mathrm{s}}{\tau_\mathrm{sp}}\right)\tensor{\mbox{\boldmath${\mu}$}}^{\pm}_\mathrm{s}({\bf B})\ .
	\label{app-22}
\end{equation}
Therefore, the total conductivity tensor $\tensor{\mbox{\boldmath${\sigma}$}}_\mathrm{tot}({\bf B})=\tensor{\mbox{\boldmath${\sigma}$}}^\|_\mathrm{c}({\bf B})+\tensor{\mbox{\boldmath${\sigma}$}}^\|_\mathrm{v}({\bf B})
+\tensor{\mbox{\boldmath${\sigma}$}}^{\pm}_\mathrm{s}({\bf B})$ can be obtained from

\begin{widetext}
\begin{multline}
\tensor{\mbox{\boldmath${\sigma}$}}_\mathrm{tot}({\bf B})=e\,\tensor{\mbox{\boldmath${\mu}$}}^\|_\mathrm{v}({\bf B})N_\mathrm{A}A_\mathrm{h}\left[(L_\mathrm{A}-W_\mathrm{p})+\int^{W_\mathrm{p}}_\mathrm{0} dz\,\exp\left(-\frac{\beta e\bar{\mu}_\mathrm{h}N_\mathrm{A}}{2\epsilon_\mathrm{0}\epsilon_\mathrm{r}D_\mathrm{h}}\,z^2\right)\right]\\
	-e\,\tensor{\mbox{\boldmath${\mu}$}}^\|_\mathrm{c}({\bf B})N_\mathrm{D}A_\mathrm{e}
\left[(L_\mathrm{D}-W_\mathrm{n})+\int^{W_\mathrm{n}}_\mathrm{0} dz\,\exp\left(-\frac{\beta e\bar{\mu}_\mathrm{e}N_\mathrm{D}}{2\epsilon_\mathrm{0}\epsilon_\mathrm{r}D_\mathrm{e}}\,z^2\right)\right]+e\,\tensor{\mbox{\boldmath${\mu}$}}^{\pm}_\mathrm{s}({\bf B})\,\left(\frac{\alpha_\mathrm{0}^2}{4\pi\hbar^2v_\mathrm{F}^2}\right)\left(L_\mathrm{A}-L_\mathrm{0}\right)^2A_\mathrm{s}\ ,
	\label{app-23}
\end{multline}
\end{widetext}
where $\alpha_\mathrm{0}$ and $L_\mathrm{0}$ are constants to be determined experimentally, $N_\mathrm{D,A}$ are doping concentrations,
$W_\mathrm{n}$ and $W_\mathrm{p}$ are depletion ranges for donors and acceptors in a $p$-$n$ junction,
$\bar{\mu}_\mathrm{e,h}$ are $\mu_\mathrm{0}(z)$ evaluated at $n_\mathrm{e,h}(z)=N_\mathrm{D,A}$, $D_\mathrm{e,h}$ are diffusion coefficients,
and $\beta=4/3$ ($\beta=7/3$) for longitudinal (Hall) conductivity. In addition,
the averaged mobilities $\tensor{\mbox{\boldmath${\mu}$}}^\|_\mathrm{c,v}({\bf B})$ are defined by
their values of $\tau_\mathrm{p(e,h)}(z)$ at $n_\mathrm{e,h}(z)=N_\mathrm{D,A}$, and three introduced coefficients are $A_s=\tau_\mathrm{s}/\tau_\mathrm{sp}\approx 3/4$,

\begin{multline}
A_\mathrm{e,h}=\left.\frac{\tau_\mathrm{e,h}(z)}{\tau_\mathrm{p(e,h)}(z)}\right|_{n_\mathrm{e,h}(z)=
N_\mathrm{D,A}}\\
	=\frac{1}{6}\left(\frac{Q_\mathrm{c}}{k_\mathrm{F}^\mathrm{e,h}}\right)^2\left[2\ln\left(\frac{2k_\mathrm{F}^\mathrm{e,h}}{Q_\mathrm{c}}\right)-1\right]\\
	=\frac{Q^2_\mathrm{c}}{6(3\pi^2N_\mathrm{D,A})^{2/3}}\left\{2\ln\left[\frac{2(3\pi^2N_\mathrm{D,A})^{1/3}}{Q_\mathrm{c}}\right]-1\right\}\ ,
	\label{app-24}
\end{multline}
where $1/Q_\mathrm{c}$ is the Thomas-Fermi screening length.
\medskip

In addition, the bulk energy-relaxation times $\tau_\mathrm{e,h}(z)$ are calculated as\,\cite{ref-1,ref-2,Huang:2004}

\begin{multline}
\frac{1}{\tau_\mathrm{e,h}(z)}=
\left[\frac{2n_\mathrm{i}}{n_\mathrm{e,h}(z)\pi\hbar Q_\mathrm{c}^2}\right]
\left(\frac{e^2}{\epsilon_\mathrm{0}\epsilon_\mathrm{r}}\right)^2\\
\times\int_\mathrm{0}^{k^\mathrm{e,h}_F(z)} dk\,{\cal D}_\mathrm{c,v}(\varepsilon^\mathrm{c,v}_\mathrm{k})\left(\frac{4k^2}{4k^2+Q_\mathrm{c}^2}\right)\\
=\left[\frac{n_\mathrm{i}m^*_\mathrm{e,h}}{8n_\mathrm{e,h}(z)\pi^3\hbar^3Q_\mathrm{c}^2}\right]\left(\frac{e^2}{\epsilon_\mathrm{0}\epsilon_\mathrm{r}}\right)^2\\
\times\left\{[2k_\mathrm{F}^\mathrm{e,h}(z)]^2-Q_\mathrm{c}^2\ln\left(\frac{[2k_\mathrm{F}^\mathrm{e,h}(z)]^2+Q_\mathrm{c}^2}{Q_\mathrm{c}^2}\right)\right\}\ ,
\label{app-25}
\end{multline}
and the surface energy-relaxation time $\tau_\mathrm{s}$ is found to be\,\cite{ref-1,ref-2,Huang:2004}

\begin{multline}
\frac{1}{\tau_\mathrm{s}}=
\frac{2\sigma_\mathrm{i}}{\pi^2\sigma_\mathrm{s}\hbar^2v_\mathrm{F}}\left(\frac{e^2}{2\epsilon_\mathrm{0}\epsilon_\mathrm{r}}\right)^2\\
\times\int_\mathrm{0}^{\pi} d\phi\,\int_\mathrm{0}^{k_\mathrm{F}^\mathrm{s}}\,\frac{k^2_\|\,dk_\|}{(q_\mathrm{c}+2k_\||\cos\phi|)^2}\ ,
\label{app-26}
\end{multline}
where $n_\mathrm{i}$ and $\sigma_\mathrm{i}$ are the impurity concentration and surface density, respectively.
\medskip

Finally, the bulk chemical potentials for electrons [$u_\mathrm{c}(z)$] and holes [$u_\mathrm{v}(z)$] are calculated as

\begin{equation}
\left[u_\mathrm{c,v}(z)\right]^{3/2}=3\pi^2\left(\frac{h^2}{2m^\ast_\mathrm{e,h}}\right)^{3/2}n_\mathrm{e,h}(z)\ ,
\label{app-27}
\end{equation}
and the carrier density functions are

\begin{multline}
n_\mathrm{e,h}(z)=N_\mathrm{D,A}\times\\
\exp\left\{-\gamma_\mathrm{e,h}\left(\frac{\bar{\mu}_\mathrm{e,h}}{D_\mathrm{e,h}}\right)\left[\Phi(z)+\gamma_\mathrm{e,h}(E_\mathrm{F}^\mathrm{e,h}/e)\right]\right\}\ .
\label{app-28}
\end{multline}
Here, the expression for the introduced potential function $\Phi(z)$ is given by

\begin{multline}
\Phi(z)=\\
\left\{\begin{array}{cc}
-E_\mathrm{F}^h/e\ , & z<-W_\mathrm{p}\\
-E_\mathrm{F}^h/e+(eN_\mathrm{A}/2\epsilon_\mathrm{0}\epsilon_\mathrm{r})\,(z+W_\mathrm{p})^2\ , & -W_\mathrm{p}<z<0\\
E_\mathrm{F}^e/e-(eN_\mathrm{D}/2\epsilon_\mathrm{0}\epsilon_\mathrm{r})\,(W_\mathrm{n}-z)^2\ , & 0<z<W_\mathrm{n}\\
E_\mathrm{F}^e/e\ , & z>W_\mathrm{n}
\end{array}\right.\ ,
\label{app-29}
\end{multline}
and $E_\mathrm{F}^\mathrm{e}$ ($E_\mathrm{F}^\mathrm{h}$) is the Fermi energy of electrons (holes) at zero temperature and defined far away from the depletion region.

\end{document}